\documentclass[ALICE,manyauthors]{cernphprep}

\usepackage[comma,square,numbers,sort&compress]{natbib}
\usepackage{hyperref}
\usepackage{lineno}

\begin{document}%

\begin{titlepage}
\PHyear{2016}
\PHnumber{102}      
\PHdate{20 April}  
%

\title{Correlated event-by-event fluctuations of flow harmonics in Pb--Pb collisions at $\sqrt{s_{_{\rm NN}}}=2.76$~TeV}
\ShortTitle{Correlated event-by-event fluctuations of flow harmonics...} 

\Collaboration{ALICE Collaboration\thanks{See Appendix~\ref{app:collab} for the list of collaboration members}}
\ShortAuthor{ALICE Collaboration} 

\begin{abstract}
We report the measurements of correlations between event-by-event fluctuations of amplitudes of anisotropic flow harmonics in nucleus--nucleus collisions, obtained for the first time using a new analysis method based on multiparticle cumulants in mixed harmonics. This novel method is robust against systematic biases originating from non-flow effects and by construction any dependence on symmetry planes is eliminated. We demonstrate that correlations of flow harmonics exhibit a better sensitivity to medium properties than the individual flow harmonics. The new measurements are performed in Pb--Pb collisions at the centre-of-mass energy per nucleon pair of $\sqrt{s_{_{\rm NN}}}=2.76$~TeV by the ALICE experiment at the Large Hadron Collider~(LHC). The centrality dependence of correlation between event-by-event fluctuations of the elliptic, $v_2$, and quadrangular, $v_4$, flow harmonics, as well as of anti-correlation between $v_2$ and triangular, $v_3$, flow harmonics are presented. The results cover two different regimes of the initial state configurations: geometry-dominated (in mid-central collisions) and fluctuation-dominated (in the most central collisions). Comparisons are made to predictions from MC-Glauber, viscous hydrodynamics, AMPT and HIJING models. Together with the existing measurements of individual flow harmonics the presented results provide further constraints on initial conditions and the transport properties of the system produced in heavy-ion collisions.
\end{abstract}
\end{titlepage}
\setcounter{page}{2}


The properties of an extreme state of matter, the Quark-Gluon Plasma (QGP), are studied by colliding heavy ions at BNL's Relativistic Heavy Ion Collider~(RHIC) and at CERN's Large Hadron Collider~(LHC). One of the most widely utilized physical phenomena in the exploration of QGP properties is collective anisotropic flow~\cite{Ollitrault:1992bk,Voloshin:2008dg}. The large elliptic flow discovered at RHIC energies~\cite{Ackermann:2000tr}, which at the LHC energy of 2.76 TeV is 30\% larger~\cite{Aamodt:2010pa} and recently reported in~\cite{Adam:2016izf} to increase even further at 5.02 TeV, demonstrated that the QGP behaves like a strongly coupled liquid with a very small ratio of the shear viscosity to entropy density $(\eta/s)$, which is close to a universal lower bound of $1/4\pi$~\cite{Kovtun:2004de}.


Anisotropic flow is traditionally quantified with harmonics $v_n$ and corresponding symmetry plane angles $\psi_n$ in the Fourier series decomposition of particle azimuthal distribution (parameterized with azimuthal angle $\varphi$) in the plane transverse to the beam direction~\cite{Voloshin:1994mz}:
\begin{equation}
\frac{dN}{d\varphi} \propto 1 + 2\sum_{n=1}^{\infty} v_{n} \cos [n(\varphi-\psi_{n})]\,.
\label{eq:Fourier}
\end{equation}
The shape of the intersecting zone of two identical heavy ions in non-central collisions is approximately ellipsoidal. This initial anisotropy is transferred via interactions among constituents and the pressure gradients developed in the QGP medium to an observable final-state anisotropic emission of particles with respect to the symmetry plane(s) of the intersecting zone. The resulting anisotropic flow for such an idealized ellipsoidal geometry is determined solely by even Fourier harmonics $v_{2n}$, and only one symmetry plane (the reaction plane) exists. Recently the importance of flow fluctuations and related additional observables have been identified. This has led to new concepts such as: non-vanishing odd harmonics $v_{2n-1}$ at midrapidity~\cite{Alver:2010gr}, non-identical symmetry plane angles $\psi_n$ and their inter-correlations~\cite{Qin:2010pf,Teaney:2010vd,Qiu:2011iv,ALICE:2011ab,Adare:2011tg,Aad:2014fla}, stochastic nature of harmonic $v_n$ and its probability density function  $P(v_n)$~\cite{Voloshin:2007pc,Gale:2012in,Aad:2013xma,Yan:2013laa,Yan:2014afa,Zhou:2015eya}, and, finally, the importance of higher order flow moments $\left<v_n^k\right>$ (where the angular brackets denote an average over all events, and $k\ge 2$)~\cite{Bhalerao:2014xra}. Two distinct regimes for anisotropic flow development are nowadays scrutinized separately: geometry-dominated (in mid-central collisions) and fluctuation-dominated (in the most central collisions)~\cite{Qiu:2011iv}.


Anisotropic flow is generated by the initial anisotropic geometry and its fluctuations coupled with an expansion of the produced medium. The initial coordinate space anisotropy can be quantified in terms of eccentricity coefficients $\varepsilon_n$ and corresponding symmetry plane angles $\Phi_n$~\cite{Alver:2010gr,Voloshin:2007pc,Alver:2006wh}. A great deal of effort is being invested to understand the relations between momentum space Fourier harmonics $v_n$ and symmetry planes $\psi_n$ on one side, and their spatial counterparts, $\varepsilon_n$ and $\Phi_n$, on the other side. These relations describe the response of the produced system to the initial coordinate space anisotropies, and therefore provide a rich repository of constraints for the system properties. In the early studies it was regularly assumed that, for small eccentricities, the harmonics $v_n$ respond linearly to the eccentricities $\varepsilon_n$ of the same order, $v_n\propto\varepsilon_n$, and that $\psi_n \simeq \Phi_n$~\cite{Alver:2010gr,Teaney:2010vd,Alver:2010dn,Lacey:2013eia}. However, for sizable eccentricities recent studies argue that the anisotropies in momentum and coordinate space are related instead with the matrix equation connecting a set of anisotropic flow harmonics $\{v_n\}$ and a set of eccentricity coefficients $\{\varepsilon_n\}$; it was demonstrated that the hydrodynamic response is both non-diagonal and non-linear, and that in general $\psi_n \neq \Phi_n$~\cite{Qin:2010pf,Qiu:2011iv,Niemi:2012aj,Yan:2015jma}. The first realization led to the conclusion that a relationship between event-by-event fluctuations of amplitudes of two different flow harmonics $v_m$ and $v_n$ can exist. This is hardly surprising for even flow harmonics in non-central collisions because the ellipsoidal shape generates non-vanishing values for all even harmonics $v_{2n}$~\cite{Kolb:2003zi}, not only for elliptic flow. However, this simple geometrical argument cannot explain the possible relation between even and odd flow harmonics in non-central collisions, and the argument is not applicable in the central collisions, where all initial shapes are equally probable since they originate solely from fluctuations. Recently a linear correlation coefficient $c(a,b)$ was defined in this context, which becomes 1 (-1) if observables $a$ and $b$ are fully linearly (antilinearly) correlated and zero in the absence of correlation~\cite{Niemi:2012aj}. Model calculations of this new observable showed that neither $v_2$ and $v_3$ nor $v_3$ and $v_4$ are linearly correlated in non-central collisions. Most importantly, it was demonstrated that $c(v_2,v_4)$ depends strongly both on the $\eta/s$ of the QGP and on the value of $c(\varepsilon_2,\varepsilon_4)$, which quantifies the relationship between corresponding eccentricities in the initial state~\cite{Niemi:2012aj}. Therefore it was concluded that new observables $c(v_n,v_m)$, depending on the choice of flow harmonics $v_n$ and $v_m$, are sensitive both to the fluctuations of the initial conditions and to the transport properties of the QGP, with the potential to discriminate between the two respective contributions when combined with a measurement of individual flow harmonics~\cite{Niemi:2012aj}. 


In this Letter we study the relationship between event-by-event fluctuations of magnitudes of two different flow harmonics of order $n$ and $m$ by using a recently proposed 4-particle observable~\cite{Bilandzic:2013kga}:
\begin{eqnarray}
\left<\left<\cos(m\varphi_1\!+\!n\varphi_2\!-\!m\varphi_3-\!n\varphi_4)\right>\right>_c &=& \left<\left<\cos(m\varphi_1\!+\!n\varphi_2\!-\!m\varphi_3-\!n\varphi_4)\right>\right>\nonumber\\
&&{}-\left<\left<\cos[m(\varphi_1\!-\!\varphi_2)]\right>\right>\left<\left<\cos[n(\varphi_1\!-\!\varphi_2)]\right>\right>\nonumber\\
&=&\left<v_{m}^2v_{n}^2\right>-\left<v_{m}^2\right>\left<v_{n}^2\right>\,,
\label{eq:4p_cumulant}
\end{eqnarray}
with the condition $m\neq n$ for two positive integers $m$ and $n$. We refer to these new observables as {\it Symmetric 2-harmonic 4-particle Cumulant}, and use notation SC$(m,n)$, or just SC. The double angular brackets indicate that the averaging procedure has been performed in two steps --- first over all distinct particle quadruplets in an event, and then in the second step the single-event averages were weighted with `number of combinations'. The latter for single-event average 4-particle correlations is mathematically equivalent to a unit weight for each individual quadruplet when the multiplicity differs event-by-event~\cite{Bilandzic:2012wva}. In both 2-particle correlators above all distinct particle pairs are considered in each case. The four-particle cumulant in Eq.~(\ref{eq:4p_cumulant}) is less sensitive to non-flow correlations than any 2- or 4-particle correlator on the right-hand side taken individually~\cite{Borghini:2000sa,Borghini:2001vi}. The last equality is true only in the absence of non-flow effects~\cite{Bhalerao:2011yg}. The observable in Eq.~(\ref{eq:4p_cumulant}) is zero in the absence of flow fluctuations, or if the magnitudes of harmonics $v_m$ and $v_n$ are uncorrelated~\cite{Bilandzic:2013kga}. It is also unaffected by relationship between symmetry plane angles $\psi_m$ and $\psi_n$. The four-particle cumulant in Eq.~(\ref{eq:4p_cumulant}) is proportional to the linear correlation coefficient $c(a,b)$ introduced in~\cite{Niemi:2012aj} and discussed above, with $a=v_m^2$ and $b=v_n^2$. Experimentally it is more reliable to measure the higher order moments of flow harmonics $v_n^k\ (k \ge 2)$ with 2- and multiparticle correlation techniques~\cite{Borghini:2001vi,Bilandzic:2010jr,Wang:1991qh}, than to measure the first moments $v_n$ with the event plane method, due to systematic uncertainties involved in the event-by-event estimation of symmetry planes~\cite{Poskanzer:1998yz,Luzum:2012da}. Therefore, we have used the new multiparticle observable in Eq.~(\ref{eq:4p_cumulant}) as meant to be the least biased measure of the correlation between event-by-event fluctuations of magnitudes of two different harmonics $v_m$ and $v_n$~\cite{Bilandzic:2013kga}.


The 2- and 4-particle correlations in Eq.~(\ref{eq:4p_cumulant}) were evaluated in terms of $Q$-vectors~\cite{Bilandzic:2010jr}. The $Q$-vector (or flow vector) in harmonic $n$ for a set of $M$ particles, where throughout this paper $M$ is multiplicity of an event, is defined as $Q_n\equiv\sum_{k=1}^Me^{in\varphi_k}$~\cite{Voloshin:1994mz,Barrette:1994xr}. We have used for a single-event average 2-particle correlation, $\left<\cos(n(\varphi_1\!-\!\varphi_2))\right>$, the following definition and analytic result in terms of $Q$-vectors:
\begin{equation}
\frac{1}{\binom{M}{2}2!}\,\sum_{\begin{subarray}{c}i,j=1\\ (i\neq j)\end{subarray}}^{M} e^{in(\varphi_i-\varphi_j)}
=\frac{1}{\binom{M}{2}2!}
\big[\left|Q_{n}\right|^2\!-\!M\big]\,.
\label{eq:two_n_n}
\end{equation}
For 4-particle correlation, $\left<\cos(m\varphi_1\!+\!n\varphi_2\!-\!m\varphi_3-\!n\varphi_4)\right>$, we used: 
\begin{eqnarray}
&&\!\!\!\!\!\! \frac{1}{\binom{M}{4}4!}\,\sum_{\begin{subarray}{c}i,j,k,l=1\\ (i\neq j\neq k\neq l)\end{subarray}}^{M} e^{i(m\varphi_i+n\varphi_j-m\varphi_k-n\varphi_l)} = \nonumber\\
&&\!\!\!\!\!\!\frac{1}{\binom{M}{4}4!}\big[\left|Q_{m}\right|^2\left|Q_{n}\right|^2\!-\!
2\mathfrak{Re}\left[Q_{m+n}Q_{m}^*Q_{n}^*\right]\!-\!
2\mathfrak{Re}\left[Q_{m}Q_{m-n}^*Q_{n}^*\right]\nonumber\\
&&{}\!\!\!\!\!\!\!+\!\left|Q_{m+n}\right|^2\!+\!\left|Q_{m-n}\right|^2\!-\!(M\!-\!4)(\left|Q_{m}\right|^2\!+\!\left|Q_{n}\right|^2)
+\!M(M\!-\!6)\big]\,.
\label{eq:four_m_n_m_m}
\end{eqnarray}
\noindent In order to obtain the all-event average correlations, denoted by $\left<\left<\cdots\right>\right>$ in Eq.~(\ref{eq:4p_cumulant}), we have weighted single-event expressions in Eqs.~(\ref{eq:two_n_n}) and (\ref{eq:four_m_n_m_m}) with weights $M(M\!-\!1)$ and $M(M\!-\!1)(M\!-\!2)(M\!-\!3)$, respectively~\cite{Bilandzic:2012wva}.


The dataset used in this analysis was obtained with the ALICE detector~\cite{Aamodt:2008zz,Abelev:2014ffa}. It consists of minimum-bias Pb--Pb collisions recorded during 2010 LHC Pb--Pb run at $\sqrt{s_{\rm NN}}=2.76$ TeV. With the default event and track selection criteria described below, we have obtained in total about $1.8\times 10^5$ events per 1\% centrality bin width. All individual systematic variations were combined in quadrature to obtain the final uncertainty.


The centrality was determined with the V0 detector~\cite{Cortese:2004aa,Abbas:2013taa,Abelev:2013qoq}. As a part of systematic checks centrality was determined independently with the Time Projection Chamber (TPC)~\cite{Alme:2010ke} and the Silicon Pixel Detector (SPD)~\cite{Dellacasa:1999kf,Aamodt:2010aa}, which have slightly worse resolution~\cite{Abelev:2013qoq}. A systematic difference of up to 3\% was observed in SC$(m,n)$ results when using different centrality estimations. Charged particles were reconstructed with the TPC and the Inner Tracking System (ITS)~\cite{Dellacasa:1999kf,Aamodt:2010aa} immersed in a 0.5 T solenoidal field. The TPC is capable of detecting charged particles in the transverse momentum range $0.1 <\!p_{\rm T}\!< 100\ {\rm GeV}/c$, with a $p_{\rm T}$ resolution of less than 6\% for tracks below $20\ {\rm GeV}/c$. Due to TPC dead zones between neighboring sectors, the track finding efficiency is about 75\% for $p_{\rm T} = 200$~MeV/$c$ and then it saturates at about 85\% for $p_{\rm T} > 1\ {\rm GeV}/c$ in Pb--Pb collisions. The TPC covers full azimuth and has a pseudorapidity coverage of $|\eta|<0.9$. Tracks reconstructed using the TPC and ITS are referred to as \emph{global}, while  tracks reconstructed only with the TPC are referred to as \emph{TPC-only}.
 

For online triggering, the V0 and SPD detectors were used~\cite{Abelev:2014ffa}. The reconstructed primary vertex is required to lie within $\pm 10$ cm of the nominal interaction point in the longitudinal direction along the beam axis. The cut on the position of the primary vertex along the beam axis was varied from $\pm 12$ cm to $\pm 6$ cm, the resulting SC measurements are consistent with those obtained with the default cut.


The main analysis was performed with global tracks selected in the transverse momentum interval $0.2\!<\!p_{\rm T}\!<\!5.0$ GeV/$c$ and pseudorapidity region $|\eta|\!<\!0.8$. With this choice of low $p_{\rm T}$ cut-off we are reducing event-by-event biases from smaller reconstruction efficiency at lower $p_{\rm T}$, while the high $p_{\rm T}$ cut-off was introduced to reduce the contribution to the anisotropies from jets. Reconstructed tracks were required to have at least 70 TPC space points (out of a maximum of 159). Only tracks with a transverse distance of closest approach (DCA) to the primary vertex less than 3 mm are accepted to reduce the contamination from secondary tracks. Tracks with kinks (the tracks that appear to change direction due to multiple scattering, $K^{\pm}$ decays) were rejected.


An independent analysis was performed with TPC-only and hybrid tracks (see below). For TPC-only tracks, the DCA cut was relaxed to 3 cm, providing different sensitivity to contamination from secondary tracks. Both the azimuthal acceptance and the reconstruction efficiency as a function of transverse momentum differ between TPC-only and global tracks. The resulting difference between independent analyses with global and TPC-only tracks was found to be 1--5\% in all the centrality ranges studied, both for SC$(3,2)$ and SC$(4,2)$. In another independent analysis with hybrid tracks, three different types of tracks were combined, in order to overcome the non-uniform azimuthal acceptance due to dead zones in SPD, and to achieve the best transverse momentum resolution~\cite{Abelev:2014ffa}. In this analysis the DCA cut was set to 3.2 cm in the longitudinal and to 2.4 cm in the transverse direction. The results between global and hybrid tracks differ by 3 to 5\%, depending on the observable considered.


One of the largest contributions to the systematic uncertainty originates from the non-uniform reconstruction efficiency as a function of transverse momentum. For the observables SC$(3,2)$ and SC$(4,2)$ the uncertainty is $7\%$ and $8\%$, respectively. In order to correct the measurements of these azimuthal correlators for various detector inefficiencies, we have constructed particle weights as a function of azimuthal angle $\varphi$ and transverse momentum $p_{\rm T}$, and used the prescription outlined in~\cite{Bilandzic:2013kga}. In particular, $p_{\rm T}$-weights were constructed as a ratio of transverse momentum distribution obtained from Monte Carlo generated tracks and from tracks reconstructed after they have passed through the detector simulated with GEANT3~\cite{Brun:1994aa}.


We have used four Monte Carlo models in this paper. The HIJING model~\cite{Wang:1991hta,Gyulassy:1994ew} was utilized to obtain the $p_{\rm T}$-weights~\cite{Bilandzic:2013kga}. Secondly, the HIJING model was used to estimate the strength of non-flow correlations (typically few-particle correlations insensitive to the collision geometry). We have evaluated the observables of interest in coordinate space by modeling the initial conditions with a MC-Glauber model~\cite{Miller:2007ri}. We have compared the centrality dependence of our observables with theoretical model from~\cite{Niemi:2015qia}, where the initial energy density profiles are calculated using a next-to-leading order perturbative-QCD+saturation model~\cite{Paatelainen:2012at,Paatelainen:2013eea}. The subsequent spacetime evolution is described by relativistic dissipative fluid dynamics with different parametrizations for the temperature dependence of the shear viscosity to entropy density ratio $\eta/s(T)$. Each of the $\eta/s(T)$ parametrizations is adjusted to reproduce the measured $v_n$ from central to mid-peripheral collisions. Finally, we provide an independent estimate of the centrality dependence of our observables by utilizing the AMPT model~\cite{Lin:2004en}.

The centrality dependence of SC$(4,2)$ (red squares) and SC$(3,2)$ (blue circles) are presented in Fig.~1. Positive values of SC$(4,2)$ are observed for all centralities. This suggests a correlation between the event-by-event fluctuations of $v_{2}$ and $v_{4}$, which indicates that finding $v_{2}$ larger than $\langle v_{2} \rangle$ in an event enhances the probability of finding $v_{4}$ larger than $\langle v_{4} \rangle$ in that event. On the other hand, the negative results of SC$(3,2)$ show the anti-correlation between $v_{2}$ and $v_{3}$ magnitudes, which further imply that finding $v_{2}$ larger than $\langle v_{2} \rangle$ enhances the probability of finding $v_{3}$ smaller than $\langle v_{3} \rangle$.
\begin{figure}[!h]
  \begin{center}
\includegraphics[keepaspectratio,width=0.65\columnwidth]{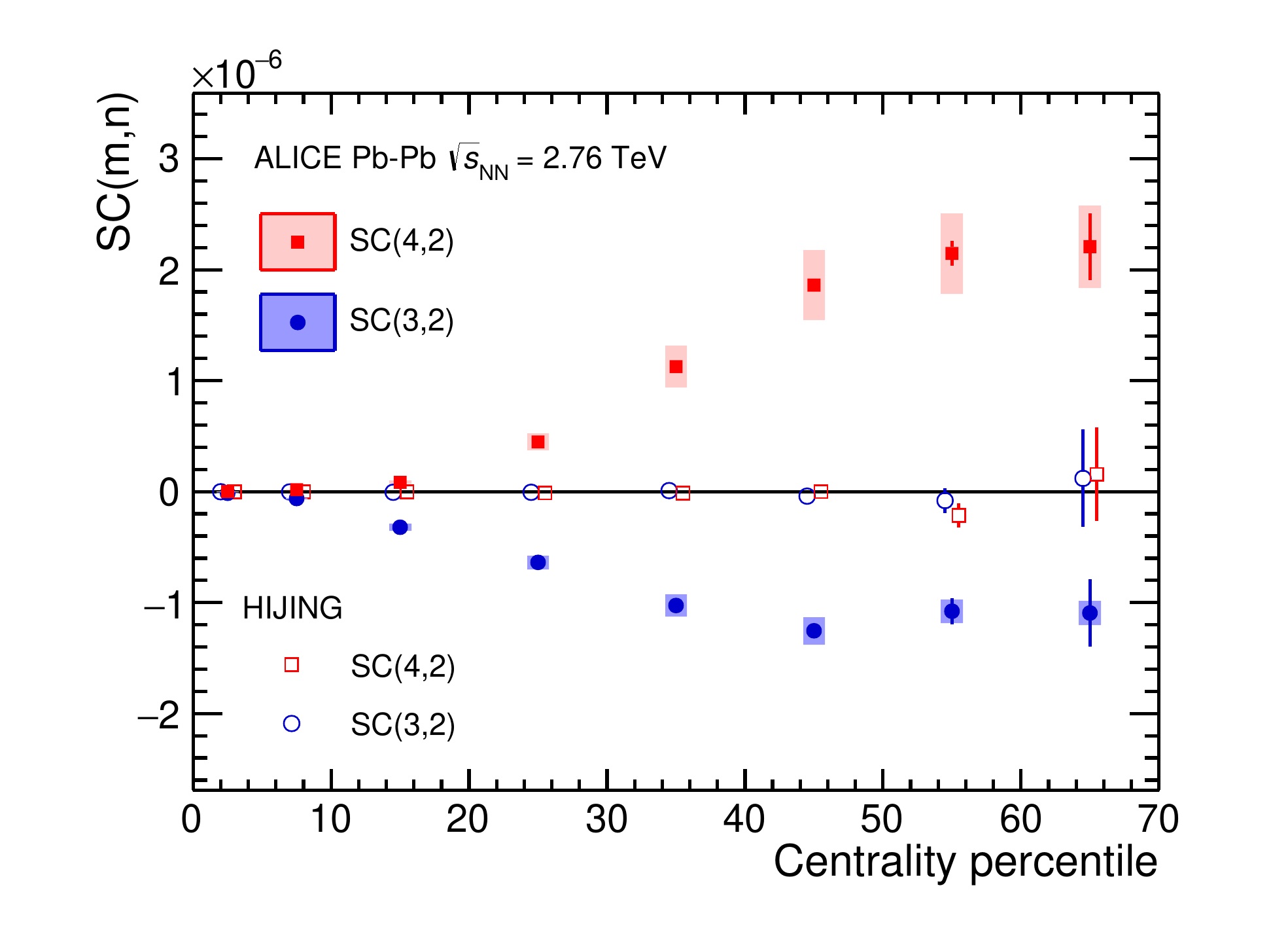} 
    \caption{Centrality dependence of observables SC$(4,2)$ (red filled squares) and SC$(3,2)$ (blue filled circles) in Pb--Pb collisions at 2.76~TeV. Systematical errors are represented with boxes. Results for HIJING model are shown with hollow markers.}
    \label{fig:Figure_1}
  \end{center}
\end{figure}
We have calculated the SC observables using HIJING which does not include anisotropic collectivity but e.g. azimuthal correlations due to jet production~\cite{Wang:1991hta,Gyulassy:1994ew}. It is found that in HIJING both $\left<\left<\cos(m\varphi_1\!+\!n\varphi_2\!-\!m\varphi_3-\!n\varphi_4)\right>\right>$ and $\left<\left<\cos[m(\varphi_1\!-\!\varphi_2)]\right>\right>\left<\left<\cos[n(\varphi_1\!-\!\varphi_2)]\right>\right>$ are non-zero. However, the calculation of SC observables from HIJING are compatible with zero for all centralities, which suggests that the SC measurements are nearly insensitive to non-flow correlations. We have also performed a study using the like-sign technique, which is another powerful approach to estimate the non-flow effects~\cite{Aamodt:2010pa}. It was found that the difference between correlations for like-sign and all charged combinations are within 10\%. This demonstrates that non-zero values of SC measurements cannot be explained by non-flow effects.

%
\begin{figure}[!h]
  \begin{center}
\includegraphics[keepaspectratio,width=0.63\columnwidth]{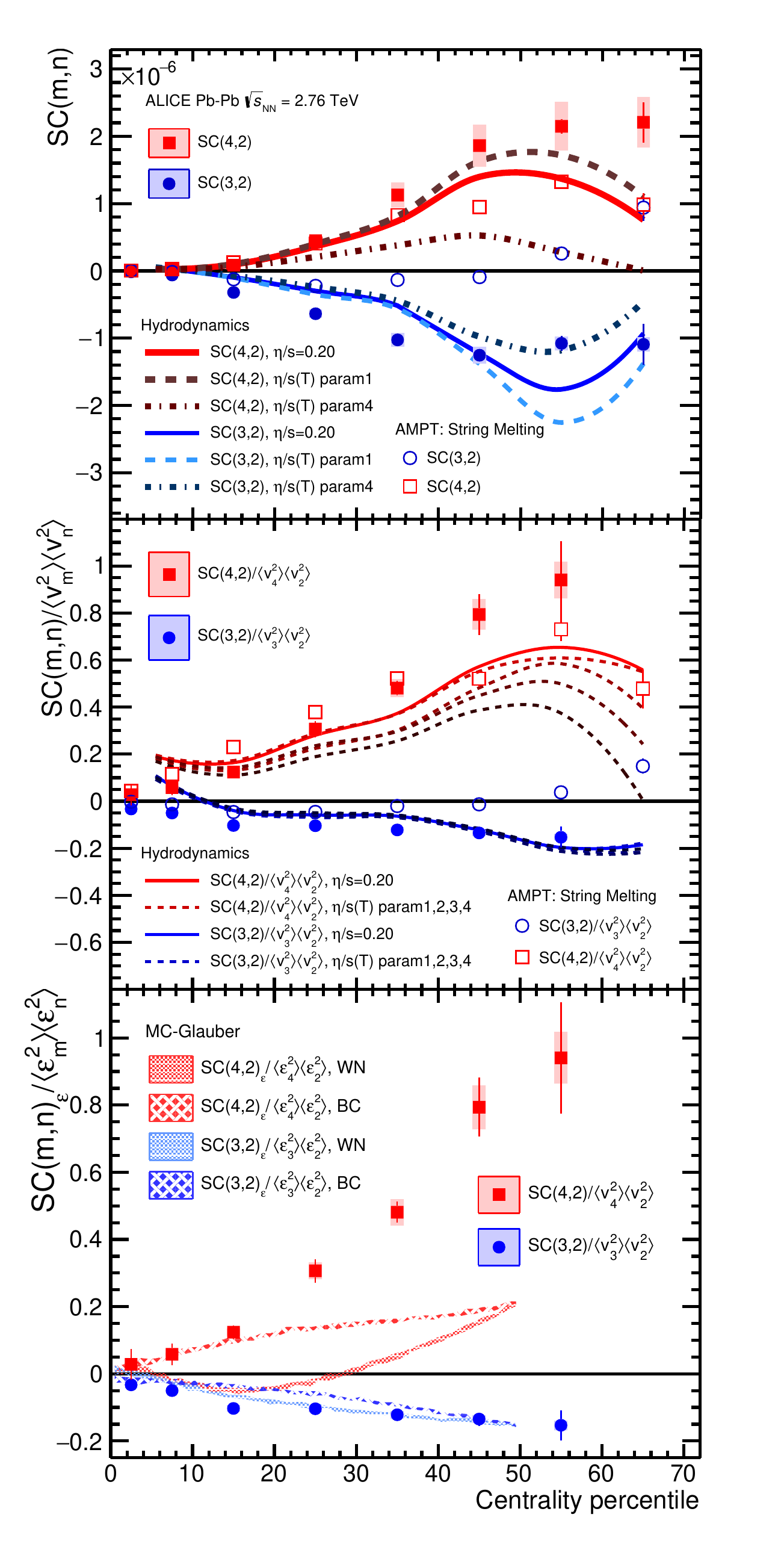} 
    \caption{AMPT model predictions are shown as hollow symbols in the (top) and (middle) panels. (top) Comparison of observables SC(4,2) (red filled squares) and SC(3,2) (blue filled circles) to theoretical model from~\cite{Niemi:2015qia}. Solid lines indicate the predictions with constant $\eta/s$, while the dashed lines indicate predictions for different parameterizations of $\eta/s$ temperature dependence (labeled in the same way as in Fig.~1 in~\cite{Niemi:2015qia}). (middle) Results divided by $\left<v_m^2\right>\left<v_n^2\right>$. (bottom) Comparison to MC-Glauber using wounded nucleon (WN) and binary collisions (BC) weights.}
    \label{fig:Figure_2}
  \end{center}
\end{figure}
A study based on the AMPT model showed that the observed (anti-)correlations are also sensitive to the transport properties, e.g. the partonic and hadronic interactions~\cite{Zhou:2015eya,Bilandzic:2013kga}. Fig.~\ref{fig:Figure_2} shows the comparison of SC$(3,2)$ and SC$(4,2)$ to the AMPT calculations which generally predict the correct sign but underestimate their magnitude. The comparison between experimental data and the theoretical calculations~\cite{Niemi:2015qia}, which incorporate both the initial conditions and system evolution, is shown in Fig.~\ref{fig:Figure_2}~(top). The model captures qualitatively the centrality dependence, but not quantitatively. Most notably, there is no single centrality for which a given $\eta/s(T)$ parameterization describes simultaneously both SC$(4,2)$ and SC$(3,2)$. On the other hand, the same theoretical model captures quantitatively the centrality dependence of individual $v_2$, $v_3$ and $v_4$ harmonics with a precision better than 10\% in central and mid-central collisions~\cite{Niemi:2015qia}. We therefore conclude that individual flow harmonics $v_n$ and new SC$(m,n)$ observables together provide a better handle on the initial conditions and $\eta/s(T)$ than each of them alone. This is emphasized in Fig.~\ref{fig:Figure_2}~(middle), where SC$(3,2)$ and SC$(4,2)$ observables were divided with the products $\left<v_3^2\right>\left<v_2^2\right>$ and $\left<v_4^2\right>\left<v_2^2\right>$, respectively, in order to obtain the normalized SC observables (the result for 60--70\% is omitted due to large statistical uncertainty). These products were obtained with 2-particle correlations and using a pseudorapidity gap of $|\Delta\eta| > 1.0$ to suppress biases from few-particle non-flow correlations. We have found that the normalized SC$(4,2)$ observable exhibits much better sensitivity to different $\eta/s(T)$ parameterizations than the normalized SC$(3,2)$ observable, see Fig.~\ref{fig:Figure_2}~(middle), and than the individual flow harmonics~\cite{Niemi:2015qia}. These findings indicate that the normalized SC$(3,2)$ observable is sensitive mainly to the initial conditions, while the normalized SC$(4,2)$ observable is sensitive to both the initial conditions and the system properties, which is consistent with the prediction from~\cite{Niemi:2012aj}.


It can be seen in Fig.~1 that SC$(4,2)$ and SC$(3,2)$ increase non-linearly up to centrality 60\%. Assuming only linear response $v_n \propto \varepsilon_n$, we expect that the normalized SC$(m,n)$ evaluated in coordinate space can capture the measurement of centrality dependence of normalized SC$(m,n)$ in the momentum space. The correlations between the $n^{\mathrm{th}}$ and $m^{\mathrm{th}}$ order harmonics were estimated with calculations of $(\left< \varepsilon_{n}^2 \varepsilon_{m}^2 \right> - \left< \varepsilon_{n}^2 \right> \left< \varepsilon_{m}^2 \right>)/\left<\varepsilon_{n}^2\right>\left<\varepsilon_{m}^2\right>$, i.e. a normalized SC observable in the coordinate space, which we denote SC$(m,n)_{\varepsilon}/\left<\varepsilon_{n}^2\right>\left<\varepsilon_{m}^2\right>$. Here the $\varepsilon_{n}$ (or $\varepsilon_{m}$) is the $n^{\mathrm{th}}$ (or $m^{\mathrm{th}}$) order coordinate space anisotropy, following the definition in~\cite{Alver:2010gr}. Different scenarios of the MC-Glauber model, named wounded nucleon (WN) and binary collisions (BC) weights, have been used. An increasing trend from central to peripheral collisions with different sign has been observed in Fig.~2~(bottom) for SC$(4,2)$ and SC$(3,2)$. A dramatic deviation of SC$(4,2)$ between data and model calculation is observed for non-central collisions. This deviation increases from mid-central to peripheral, which could be understood as the contribution of the non-linear response ($\varepsilon_{2}$ contributes to $v_4$)  increasing as a function of centrality, which is consistent with that reported in~\cite{Aad:2015lwa}. Since the normalized SC$(3,2)$ appears to be sensitive only to initial conditions and not to $\eta/s(T)$, see Fig.~2~(middle), MC-Glauber model captures better its centrality dependence than for normalized SC$(4,2)$ observable, see Fig.~2~(bottom). 


The relationship between the flow harmonics $v_2$, $v_3$, $v_4$ have also been investigated by the ATLAS Collaboration using the ESE technique~\cite{Schukraft:2012ah,Petersen:2013vca,Aad:2015lwa}. For events with a larger $v_2$, the ATLAS Collaboration showed these have a smaller than average $v_3$, and a larger than average $v_4$. For events with a smaller $v_2$, the opposite trend occurred. These observations are consistent with the patterns observed via the SC measurements presented in this Letter. The SC observables, however, provide a compact quantitative measure of these correlations, without fitting correlations between $v_n$ and $v_m$. This simplifies the quantitative comparison of the SC observables with hydrodynamical calculations as shown in Fig.~\ref{fig:Figure_2}.

\begin{figure}[!h]
  \begin{center}
\includegraphics[keepaspectratio,width=0.65\columnwidth]{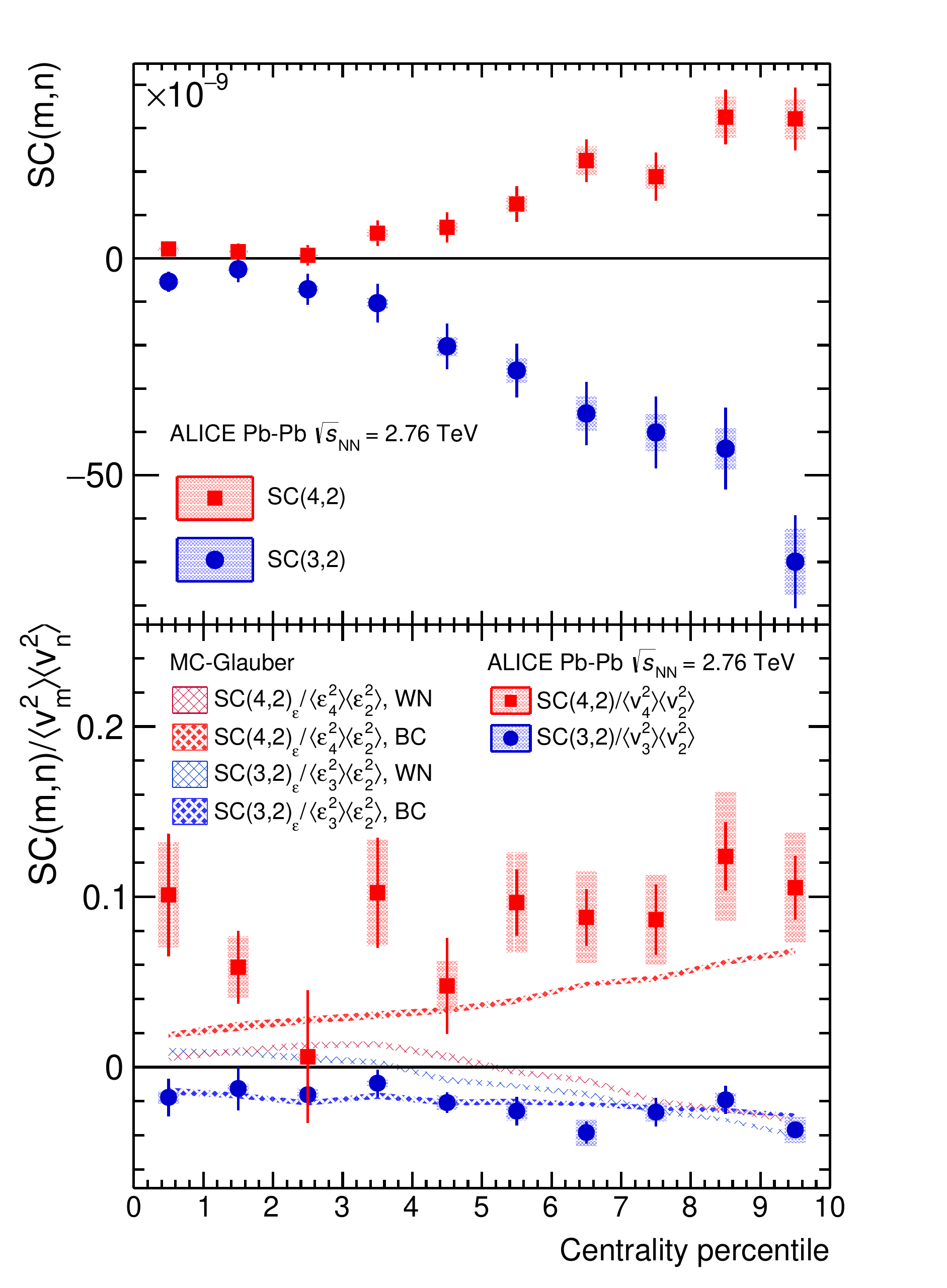} 
    \caption{Correlated and anti-correlated event-by-event fluctuations in coordinate (MC-Glauber) and momentum space (data). On the bottom panel we present the normalized SC observables, where pseudorapidity gap $|\Delta\eta| > 1.0$ was applied in both two-particle correlations in denominator used to estimate individual flow harmonics.}
    \label{fig:Figure_3}
  \end{center}
\end{figure}

In the most central collisions the anisotropies originate mainly from fluctuations, i.e.\ the initial ellipsoidal geometry characteristic for mid-central collisions plays little role in this regime. Therefore we have performed a separate analysis for centrality range 0--10\% in centrality bins of 1\%. The results are presented in Fig.~\ref{fig:Figure_3}. 
We observe that event-by-event fluctuations of $v_2$ and $v_4$ remain correlated, and of $v_2$ and $v_3$ anti-correlated, also in this regime. However, the strength of the (anti)-correlations exhibits a different centrality dependence than for the wider centrality range shown in Fig.~\ref{fig:Figure_1}. As seen in Fig.~\ref{fig:Figure_3}~(top) the centrality dependence cannot be linearly extrapolated from the 0--10\% region to the full centrality range. Comparison with two different parameterizations of the MC-Glauber initial conditions for normalized SC observables presented in Fig.~\ref{fig:Figure_3}~(bottom) suggests that the BC parameterization (binary collisions weights) is favored by the data in most central collisions. This agreement may suggest the scaling with the number of quark participants~\cite{Eremin:2003qn,Miller:2003kd,Adamczyk:2015obl,Adare:2015bua,Loizides:2016djv} in central collisions at the LHC energies.


In summary, we have measured for the first time the new multiparticle observables, the Symmetric 2-harmonic 4-particle Cumulants (SC), which quantify the relationship between event-by-event fluctuations of two different flow harmonics. We have found that fluctuations of $v_2$ and $v_3$ are anti-correlated in all centralities, however the details of the centrality dependence differ in the fluctutation-dominated (most central) and the geometry-dominated (mid-central) regimes. Fluctuations of $v_2$ and $v_4$ are correlated for all centralities. The SC observables were used to discriminate between the state-of-the-art hydro model calculations with different parameterizations of the temperature dependence of $\eta/s$, for all of which the centrality dependence of elliptic, triangular and quadrangular flow has weaker sensitivity at the LHC. In particular, the centrality dependence of SC$(4,2)$ cannot be captured with the constant $\eta/s$. We have also used our results to discriminate between two different parameterizations of initial conditions and have demonstrated that in the fluctuation-dominated regime (in central collisions) MC-Glauber initial conditions with binary collisions weights are favored over wounded nucleon weights.


\newenvironment{acknowledgement}{\relax}{\relax}
\begin{acknowledgement}
\section*{Acknowledgements}
\noindent The ALICE Collaboration would like to thank Harri Niemi for providing  the latest predictions from the state-of-the-art hydrodynamic model.

The ALICE Collaboration would like to thank all its engineers and technicians for their invaluable contributions to the construction of the experiment and the CERN accelerator teams for the outstanding performance of the LHC complex.
The ALICE Collaboration gratefully acknowledges the resources and support provided by all Grid centres and the Worldwide LHC Computing Grid (WLCG) collaboration.
The ALICE Collaboration acknowledges the following funding agencies for their support in building and
running the ALICE detector:
State Committee of Science,  World Federation of Scientists (WFS)
and Swiss Fonds Kidagan, Armenia;
Conselho Nacional de Desenvolvimento Cient\'{\i}fico e Tecnol\'{o}gico (CNPq), Financiadora de Estudos e Projetos (FINEP),
Funda\c{c}\~{a}o de Amparo \`{a} Pesquisa do Estado de S\~{a}o Paulo (FAPESP);
Ministry of Science and Technology of China (MSTC), National Natural Science Foundation of China (NSFC) and Ministry of Education of China (MOEC)";
Ministry of Science, Education and Sports of Croatia and  Unity through Knowledge Fund, Croatia;
Ministry of Education and Youth of the Czech Republic;
Danish Natural Science Research Council, the Carlsberg Foundation and the Danish National Research Foundation;
The European Research Council under the European Community's Seventh Framework Programme;
Helsinki Institute of Physics and the Academy of Finland;
French CNRS-IN2P3, the `Region Pays de Loire', `Region Alsace', `Region Auvergne' and CEA, France;
German Bundesministerium fur Bildung, Wissenschaft, Forschung und Technologie (BMBF) and the Helmholtz Association;
General Secretariat for Research and Technology, Ministry of Development, Greece;
National Research, Development and Innovation Office (NKFIH), Hungary;
Council of Scientific and Industrial Research (CSIR), New Delhi;
Department of Atomic Energy and Department of Science and Technology of the Government of India;
Istituto Nazionale di Fisica Nucleare (INFN) and Centro Fermi - Museo Storico della Fisica e Centro Studi e Ricerche ``Enrico Fermi'', Italy;
Japan Society for the Promotion of Science (JSPS) KAKENHI and MEXT, Japan;
National Research Foundation of Korea (NRF);
Consejo Nacional de Cienca y Tecnologia (CONACYT), Direccion General de Asuntos del Personal Academico(DGAPA), M\'{e}xico, Amerique Latine Formation academique - 
European Commission~(ALFA-EC) and the EPLANET Program~(European Particle Physics Latin American Network);
Stichting voor Fundamenteel Onderzoek der Materie (FOM) and the Nederlandse Organisatie voor Wetenschappelijk Onderzoek (NWO), Netherlands;
Research Council of Norway (NFR);
Pontificia Universidad Cat\'{o}lica del Per\'{u};
National Science Centre, Poland;
Ministry of National Education/Institute for Atomic Physics and National Council of Scientific Research in Higher Education~(CNCSI-UEFISCDI), Romania;
Joint Institute for Nuclear Research, Dubna;
Ministry of Education and Science of Russian Federation, Russian Academy of Sciences, Russian Federal Agency of Atomic Energy, Russian Federal Agency for Science and Innovations and The Russian Foundation for Basic Research;
Ministry of Education of Slovakia;
Department of Science and Technology, South Africa;
Centro de Investigaciones Energeticas, Medioambientales y Tecnologicas (CIEMAT), E-Infrastructure shared between Europe and Latin America (EELA), 
Ministerio de Econom\'{i}a y Competitividad (MINECO) of Spain, Xunta de Galicia (Conseller\'{\i}a de Educaci\'{o}n),
Centro de Aplicaciones Tecnológicas y Desarrollo Nuclear (CEA\-DEN), Cubaenerg\'{\i}a, Cuba, and IAEA (International Atomic Energy Agency);
Swedish Research Council (VR) and Knut $\&$ Alice Wallenberg Foundation (KAW);
National Science and Technology Development Agency (NSDTA), Suranaree University of Technology (SUT) and Office of the Higher Education Commission under NRU project of Thailand;
Ukraine Ministry of Education and Science;
United Kingdom Science and Technology Facilities Council (STFC);
The United States Department of Energy, the United States National Science Foundation, the State of Texas, and the State of Ohio.

\end{acknowledgement}

\bibliographystyle{utphys} 
\bibliography{bibliography}

\providecommand{\href}[2]{#2}\begingroup\raggedright\begin{thebibliography}{10}

\bibitem{Ollitrault:1992bk}
J.-Y. Ollitrault, ``{Anisotropy as a signature of transverse collective
  flow},''
\href{http://dx.doi.org/10.1103/PhysRevD.46.229}{{\em Phys. Rev.} {\bfseries
  D46} (1992) 229--245}.

\bibitem{Voloshin:2008dg}
S.~A. Voloshin, A.~M. Poskanzer, and R.~Snellings, ``{Collective phenomena in
  non-central nuclear collisions},''
\href{http://arxiv.org/abs/0809.2949}{{\ttfamily arXiv:0809.2949 [nucl-ex]}}.

\bibitem{Ackermann:2000tr}
{\bfseries STAR} Collaboration, K.~H. Ackermann {\em et~al.}, ``{Elliptic flow
  in Au + Au collisions at $\sqrt{s_{_{\rm NN}}}$ = 130 GeV},''
  \href{http://dx.doi.org/10.1103/PhysRevLett.86.402}{{\em Phys. Rev. Lett.}
  {\bfseries 86} (2001) 402--407},
\href{http://arxiv.org/abs/nucl-ex/0009011}{{\ttfamily arXiv:nucl-ex/0009011
  [nucl-ex]}}.

\bibitem{Aamodt:2010pa}
{\bfseries ALICE} Collaboration, K.~Aamodt {\em et~al.}, ``{Elliptic flow of
  charged particles in Pb-Pb collisions at 2.76 TeV},''
  \href{http://dx.doi.org/10.1103/PhysRevLett.105.252302}{{\em Phys. Rev.
  Lett.} {\bfseries 105} (2010) 252302},
\href{http://arxiv.org/abs/1011.3914}{{\ttfamily arXiv:1011.3914 [nucl-ex]}}.

\bibitem{Adam:2016izf}
{\bfseries ALICE} Collaboration, J.~Adam {\em et~al.}, ``{Anisotropic flow of
  charged particles in Pb-Pb collisions at $\sqrt{s_{\rm NN}}=5.02$ TeV},''
  \href{http://dx.doi.org/10.1103/PhysRevLett.116.132302}{{\em Phys. Rev.
  Lett.} {\bfseries 116} no.~13, (2016) 132302},
\href{http://arxiv.org/abs/1602.01119}{{\ttfamily arXiv:1602.01119 [nucl-ex]}}.

\bibitem{Kovtun:2004de}
P.~Kovtun, D.~T. Son, and A.~O. Starinets, ``{Viscosity in strongly interacting
  quantum field theories from black hole physics},''
  \href{http://dx.doi.org/10.1103/PhysRevLett.94.111601}{{\em Phys. Rev. Lett.}
  {\bfseries 94} (2005) 111601},
\href{http://arxiv.org/abs/hep-th/0405231}{{\ttfamily arXiv:hep-th/0405231
  [hep-th]}}.

\bibitem{Voloshin:1994mz}
S.~Voloshin and Y.~Zhang, ``{Flow study in relativistic nuclear collisions by
  Fourier expansion of Azimuthal particle distributions},''
  \href{http://dx.doi.org/10.1007/s002880050141}{{\em Z. Phys.} {\bfseries C70}
  (1996) 665--672},
\href{http://arxiv.org/abs/hep-ph/9407282}{{\ttfamily arXiv:hep-ph/9407282
  [hep-ph]}}.

\bibitem{Alver:2010gr}
B.~Alver and G.~Roland, ``{Collision geometry fluctuations and triangular flow
  in heavy-ion collisions},''
  \href{http://dx.doi.org/10.1103/PhysRevC.82.039903,
  10.1103/PhysRevC.81.054905}{{\em Phys. Rev.} {\bfseries C81} (2010) 054905},
  \href{http://arxiv.org/abs/1003.0194}{{\ttfamily arXiv:1003.0194 [nucl-th]}}.
[Erratum: Phys. Rev.C82,039903(2010)].

\bibitem{Qin:2010pf}
G.-Y. Qin, H.~Petersen, S.~A. Bass, and B.~Muller, ``{Translation of collision
  geometry fluctuations into momentum anisotropies in relativistic heavy-ion
  collisions},'' \href{http://dx.doi.org/10.1103/PhysRevC.82.064903}{{\em Phys.
  Rev.} {\bfseries C82} (2010) 064903},
\href{http://arxiv.org/abs/1009.1847}{{\ttfamily arXiv:1009.1847 [nucl-th]}}.

\bibitem{Teaney:2010vd}
D.~Teaney and L.~Yan, ``{Triangularity and Dipole Asymmetry in Heavy Ion
  Collisions},'' \href{http://dx.doi.org/10.1103/PhysRevC.83.064904}{{\em Phys.
  Rev.} {\bfseries C83} (2011) 064904},
\href{http://arxiv.org/abs/1010.1876}{{\ttfamily arXiv:1010.1876 [nucl-th]}}.

\bibitem{Qiu:2011iv}
Z.~Qiu and U.~W. Heinz, ``{Event-by-event shape and flow fluctuations of
  relativistic heavy-ion collision fireballs},''
  \href{http://dx.doi.org/10.1103/PhysRevC.84.024911}{{\em Phys. Rev.}
  {\bfseries C84} (2011) 024911},
\href{http://arxiv.org/abs/1104.0650}{{\ttfamily arXiv:1104.0650 [nucl-th]}}.

\bibitem{ALICE:2011ab}
{\bfseries ALICE} Collaboration, K.~Aamodt {\em et~al.}, ``{Higher harmonic
  anisotropic flow measurements of charged particles in Pb-Pb collisions at
  $\sqrt{s_{\rm NN}}$ = 2.76 TeV},''
  \href{http://dx.doi.org/10.1103/PhysRevLett.107.032301}{{\em Phys. Rev.
  Lett.} {\bfseries 107} (2011) 032301},
\href{http://arxiv.org/abs/1105.3865}{{\ttfamily arXiv:1105.3865 [nucl-ex]}}.

\bibitem{Adare:2011tg}
{\bfseries PHENIX} Collaboration, A.~Adare {\em et~al.}, ``{Measurements of
  Higher-Order Flow Harmonics in Au+Au Collisions at $\sqrt{s_{\rm NN}} = 200$
  GeV},'' \href{http://dx.doi.org/10.1103/PhysRevLett.107.252301}{{\em Phys.
  Rev. Lett.} {\bfseries 107} (2011) 252301},
\href{http://arxiv.org/abs/1105.3928}{{\ttfamily arXiv:1105.3928 [nucl-ex]}}.

\bibitem{Aad:2014fla}
{\bfseries ATLAS} Collaboration, G.~Aad {\em et~al.}, ``{Measurement of
  event-plane correlations in $\sqrt{s_{\rm NN}} = 2.76$ TeV lead-lead
  collisions with the ATLAS detector},''
  \href{http://dx.doi.org/10.1103/PhysRevC.90.024905}{{\em Phys. Rev.}
  {\bfseries C90} no.~2, (2014) 024905},
\href{http://arxiv.org/abs/1403.0489}{{\ttfamily arXiv:1403.0489 [hep-ex]}}.

\bibitem{Voloshin:2007pc}
S.~A. Voloshin, A.~M. Poskanzer, A.~Tang, and G.~Wang, ``{Elliptic flow in the
  Gaussian model of eccentricity fluctuations},''
  \href{http://dx.doi.org/10.1016/j.physletb.2007.11.043}{{\em Phys. Lett.}
  {\bfseries B659} (2008) 537--541},
\href{http://arxiv.org/abs/0708.0800}{{\ttfamily arXiv:0708.0800 [nucl-th]}}.

\bibitem{Gale:2012in}
C.~Gale, S.~Jeon, B.~Schenke, P.~Tribedy, and R.~Venugopalan, ``{Initial state
  fluctuations and higher harmonic flow in heavy-ion collisions},''
  \href{http://dx.doi.org/10.1016/j.nuclphysa.2013.02.037}{{\em Nucl. Phys.}
  {\bfseries A904-905} (2013) 409c--412c},
\href{http://arxiv.org/abs/1210.5144}{{\ttfamily arXiv:1210.5144 [hep-ph]}}.

\bibitem{Aad:2013xma}
{\bfseries ATLAS} Collaboration, G.~Aad {\em et~al.}, ``{Measurement of the
  distributions of event-by-event flow harmonics in lead-lead collisions at
  $\sqrt{s_{\rm NN}}$ = 2.76 TeV with the ATLAS detector at the LHC},''
  \href{http://dx.doi.org/10.1007/JHEP11(2013)183}{{\em JHEP} {\bfseries 11}
  (2013) 183},
\href{http://arxiv.org/abs/1305.2942}{{\ttfamily arXiv:1305.2942 [hep-ex]}}.

\bibitem{Yan:2013laa}
L.~Yan and J.-Y. Ollitrault, ``{Universal fluctuation-driven eccentricities in
  proton-proton, proton-nucleus and nucleus-nucleus collisions},''
  \href{http://dx.doi.org/10.1103/PhysRevLett.112.082301}{{\em Phys. Rev.
  Lett.} {\bfseries 112} (2014) 082301},
\href{http://arxiv.org/abs/1312.6555}{{\ttfamily arXiv:1312.6555 [nucl-th]}}.

\bibitem{Yan:2014afa}
L.~Yan, J.-Y. Ollitrault, and A.~M. Poskanzer, ``{Eccentricity distributions in
  nucleus-nucleus collisions},''
  \href{http://dx.doi.org/10.1103/PhysRevC.90.024903}{{\em Phys. Rev.}
  {\bfseries C90} no.~2, (2014) 024903},
\href{http://arxiv.org/abs/1405.6595}{{\ttfamily arXiv:1405.6595 [nucl-th]}}.

\bibitem{Zhou:2015eya}
Y.~Zhou, K.~Xiao, Z.~Feng, F.~Liu, and R.~Snellings, ``{Anisotropic
  distributions in a multi-phase transport model},''
\href{http://arxiv.org/abs/1508.03306}{{\ttfamily arXiv:1508.03306 [nucl-ex]}}.

\bibitem{Bhalerao:2014xra}
R.~S. Bhalerao, J.-Y. Ollitrault, and S.~Pal, ``{Characterizing flow
  fluctuations with moments},''
  \href{http://dx.doi.org/10.1016/j.physletb.2015.01.019}{{\em Phys. Lett.}
  {\bfseries B742} (2015) 94--98},
\href{http://arxiv.org/abs/1411.5160}{{\ttfamily arXiv:1411.5160 [nucl-th]}}.

\bibitem{Alver:2006wh}
{\bfseries PHOBOS} Collaboration, B.~Alver {\em et~al.}, ``{System size,
  energy, pseudorapidity, and centrality dependence of elliptic flow},''
  \href{http://dx.doi.org/10.1103/PhysRevLett.98.242302}{{\em Phys. Rev. Lett.}
  {\bfseries 98} (2007) 242302},
\href{http://arxiv.org/abs/nucl-ex/0610037}{{\ttfamily arXiv:nucl-ex/0610037
  [nucl-ex]}}.

\bibitem{Alver:2010dn}
B.~H. Alver, C.~Gombeaud, M.~Luzum, and J.-Y. Ollitrault, ``{Triangular flow in
  hydrodynamics and transport theory},''
  \href{http://dx.doi.org/10.1103/PhysRevC.82.034913}{{\em Phys. Rev.}
  {\bfseries C82} (2010) 034913},
\href{http://arxiv.org/abs/1007.5469}{{\ttfamily arXiv:1007.5469 [nucl-th]}}.

\bibitem{Lacey:2013eia}
R.~A. Lacey, D.~Reynolds, A.~Taranenko, N.~N. Ajitanand, J.~M. Alexander, F.-H.
  Liu, Y.~Gu, and A.~Mwai, ``{Acoustic scaling of anisotropic flow in
  shape-engineered events: implications for extraction of the specific shear
  viscosity of the quark gluon plasma},''
\href{http://arxiv.org/abs/1311.1728}{{\ttfamily arXiv:1311.1728 [nucl-ex]}}.

\bibitem{Niemi:2012aj}
H.~Niemi, G.~S. Denicol, H.~Holopainen, and P.~Huovinen, ``{Event-by-event
  distributions of azimuthal asymmetries in ultrarelativistic heavy-ion
  collisions},'' \href{http://dx.doi.org/10.1103/PhysRevC.87.054901}{{\em Phys.
  Rev.} {\bfseries C87} no.~5, (2013) 054901},
\href{http://arxiv.org/abs/1212.1008}{{\ttfamily arXiv:1212.1008 [nucl-th]}}.

\bibitem{Yan:2015jma}
L.~Yan and J.-Y. Ollitrault, ``{$\nu_4, \nu_5, \nu_6, \nu_7$: nonlinear
  hydrodynamic response versus LHC data},''
  \href{http://dx.doi.org/10.1016/j.physletb.2015.03.040}{{\em Phys. Lett.}
  {\bfseries B744} (2015) 82--87},
\href{http://arxiv.org/abs/1502.02502}{{\ttfamily arXiv:1502.02502 [nucl-th]}}.

\bibitem{Kolb:2003zi}
P.~F. Kolb, ``{$v_{4}$: A Small, but sensitive observable for heavy ion
  collisions},'' \href{http://dx.doi.org/10.1103/PhysRevC.68.031902}{{\em Phys.
  Rev.} {\bfseries C68} (2003) 031902},
\href{http://arxiv.org/abs/nucl-th/0306081}{{\ttfamily arXiv:nucl-th/0306081
  [nucl-th]}}.

\bibitem{Bilandzic:2013kga}
A.~Bilandzic, C.~H. Christensen, K.~Gulbrandsen, A.~Hansen, and Y.~Zhou,
  ``{Generic framework for anisotropic flow analyses with multiparticle
  azimuthal correlations},''
  \href{http://dx.doi.org/10.1103/PhysRevC.89.064904}{{\em Phys. Rev.}
  {\bfseries C89} no.~6, (2014) 064904},
\href{http://arxiv.org/abs/1312.3572}{{\ttfamily arXiv:1312.3572 [nucl-ex]}}.

\bibitem{Bilandzic:2012wva}
A.~Bilandzic, {\em {Anisotropic flow measurements in ALICE at the large hadron
  collider (see Section 3.1.4)}}.
\newblock PhD thesis, Utrecht U., 2012.
\newblock
\url{https://inspirehep.net/record/1186272/files/CERN-THESIS-2012-018.pdf}.
\newblock

\bibitem{Borghini:2000sa}
N.~Borghini, P.~M. Dinh, and J.-Y. Ollitrault, ``{A New method for measuring
  azimuthal distributions in nucleus-nucleus collisions},''
  \href{http://dx.doi.org/10.1103/PhysRevC.63.054906}{{\em Phys. Rev.}
  {\bfseries C63} (2001) 054906},
\href{http://arxiv.org/abs/nucl-th/0007063}{{\ttfamily arXiv:nucl-th/0007063
  [nucl-th]}}.

\bibitem{Borghini:2001vi}
N.~Borghini, P.~M. Dinh, and J.-Y. Ollitrault, ``{Flow analysis from
  multiparticle azimuthal correlations},''
  \href{http://dx.doi.org/10.1103/PhysRevC.64.054901}{{\em Phys. Rev.}
  {\bfseries C64} (2001) 054901},
\href{http://arxiv.org/abs/nucl-th/0105040}{{\ttfamily arXiv:nucl-th/0105040
  [nucl-th]}}.

\bibitem{Bhalerao:2011yg}
R.~S. Bhalerao, M.~Luzum, and J.-Y. Ollitrault, ``{Determining initial-state
  fluctuations from flow measurements in heavy-ion collisions},''
  \href{http://dx.doi.org/10.1103/PhysRevC.84.034910}{{\em Phys. Rev.}
  {\bfseries C84} (2011) 034910},
\href{http://arxiv.org/abs/1104.4740}{{\ttfamily arXiv:1104.4740 [nucl-th]}}.

\bibitem{Bilandzic:2010jr}
A.~Bilandzic, R.~Snellings, and S.~Voloshin, ``{Flow analysis with cumulants:
  Direct calculations},''
  \href{http://dx.doi.org/10.1103/PhysRevC.83.044913}{{\em Phys. Rev.}
  {\bfseries C83} (2011) 044913},
\href{http://arxiv.org/abs/1010.0233}{{\ttfamily arXiv:1010.0233 [nucl-ex]}}.

\bibitem{Wang:1991qh}
S.~Wang, Y.~Z. Jiang, Y.~M. Liu, D.~Keane, D.~Beavis, S.~Y. Chu, S.~Y. Fung,
  M.~Vient, C.~Hartnack, and H.~Stoecker, ``{Measurement of collective flow in
  heavy ion collisions using particle pair correlations},''
\href{http://dx.doi.org/10.1103/PhysRevC.44.1091}{{\em Phys. Rev.} {\bfseries
  C44} (1991) 1091--1095}.

\bibitem{Poskanzer:1998yz}
A.~M. Poskanzer and S.~A. Voloshin, ``{Methods for analyzing anisotropic flow
  in relativistic nuclear collisions},''
  \href{http://dx.doi.org/10.1103/PhysRevC.58.1671}{{\em Phys. Rev.} {\bfseries
  C58} (1998) 1671--1678},
\href{http://arxiv.org/abs/nucl-ex/9805001}{{\ttfamily arXiv:nucl-ex/9805001
  [nucl-ex]}}.

\bibitem{Luzum:2012da}
M.~Luzum and J.-Y. Ollitrault, ``{Eliminating experimental bias in
  anisotropic-flow measurements of high-energy nuclear collisions},''
  \href{http://dx.doi.org/10.1103/PhysRevC.87.044907}{{\em Phys. Rev.}
  {\bfseries C87} no.~4, (2013) 044907},
\href{http://arxiv.org/abs/1209.2323}{{\ttfamily arXiv:1209.2323 [nucl-ex]}}.

\bibitem{Barrette:1994xr}
{\bfseries E877} Collaboration, J.~Barrette {\em et~al.}, ``{Observation of
  anisotropic event shapes and transverse flow in Au + Au collisions at AGS
  energy},'' \href{http://dx.doi.org/10.1103/PhysRevLett.73.2532}{{\em Phys.
  Rev. Lett.} {\bfseries 73} (1994) 2532--2535},
\href{http://arxiv.org/abs/hep-ex/9405003}{{\ttfamily arXiv:hep-ex/9405003
  [hep-ex]}}.

\bibitem{Aamodt:2008zz}
{\bfseries ALICE} Collaboration, K.~Aamodt {\em et~al.}, ``{The ALICE
  experiment at the CERN LHC},''
\href{http://dx.doi.org/10.1088/1748-0221/3/08/S08002}{{\em JINST} {\bfseries
  3} (2008) S08002}.

\bibitem{Abelev:2014ffa}
{\bfseries ALICE} Collaboration, B.~B. Abelev {\em et~al.}, ``{Performance of
  the ALICE Experiment at the CERN LHC},''
  \href{http://dx.doi.org/10.1142/S0217751X14300440}{{\em Int. J. Mod. Phys.}
  {\bfseries A29} (2014) 1430044},
\href{http://arxiv.org/abs/1402.4476}{{\ttfamily arXiv:1402.4476 [nucl-ex]}}.

\bibitem{Cortese:2004aa}
{\bfseries ALICE} Collaboration, P.~Cortese {\em et~al.},
``{ALICE technical design report on forward detectors: FMD, T0 and V0,
  CERN-LHCC-2004-025},''.

\bibitem{Abbas:2013taa}
{\bfseries ALICE} Collaboration, E.~Abbas {\em et~al.}, ``{Performance of the
  ALICE VZERO system},''
  \href{http://dx.doi.org/10.1088/1748-0221/8/10/P10016}{{\em JINST} {\bfseries
  8} (2013) P10016},
\href{http://arxiv.org/abs/1306.3130}{{\ttfamily arXiv:1306.3130 [nucl-ex]}}.

\bibitem{Abelev:2013qoq}
{\bfseries ALICE} Collaboration, B.~Abelev {\em et~al.}, ``{Centrality
  determination of Pb-Pb collisions at $\sqrt{s_{\rm NN}}$ = 2.76 TeV with
  ALICE},'' \href{http://dx.doi.org/10.1103/PhysRevC.88.044909}{{\em Phys.
  Rev.} {\bfseries C88} no.~4, (2013) 044909},
\href{http://arxiv.org/abs/1301.4361}{{\ttfamily arXiv:1301.4361 [nucl-ex]}}.

\bibitem{Alme:2010ke}
J.~Alme {\em et~al.}, ``{The ALICE TPC, a large 3-dimensional tracking device
  with fast readout for ultra-high multiplicity events},''
  \href{http://dx.doi.org/10.1016/j.nima.2010.04.042}{{\em Nucl. Instrum.
  Meth.} {\bfseries A622} (2010) 316--367},
\href{http://arxiv.org/abs/1001.1950}{{\ttfamily arXiv:1001.1950
  [physics.ins-det]}}.

\bibitem{Dellacasa:1999kf}
{\bfseries ALICE} Collaboration, G.~Dellacasa {\em et~al.},
``{ALICE technical design report of the inner tracking system (ITS),
  CERN-LHCC-99-12},''.

\bibitem{Aamodt:2010aa}
{\bfseries ALICE} Collaboration, K.~Aamodt {\em et~al.}, ``{Alignment of the
  ALICE Inner Tracking System with cosmic-ray tracks},''
  \href{http://dx.doi.org/10.1088/1748-0221/5/03/P03003}{{\em JINST} {\bfseries
  5} (2010) P03003},
\href{http://arxiv.org/abs/1001.0502}{{\ttfamily arXiv:1001.0502
  [physics.ins-det]}}.

\bibitem{Brun:1994aa}
R.~Brun, F.~Carminati, and S.~Giani,
``{GEANT Detector Description and Simulation Tool, CERN-W5013, 1994},''.

\bibitem{Wang:1991hta}
X.-N. Wang and M.~Gyulassy, ``{HIJING: A Monte Carlo model for multiple jet
  production in pp, pA and AA collisions},''
\href{http://dx.doi.org/10.1103/PhysRevD.44.3501}{{\em Phys. Rev.} {\bfseries
  D44} (1991) 3501--3516}.

\bibitem{Gyulassy:1994ew}
M.~Gyulassy and X.-N. Wang, ``{HIJING 1.0: A Monte Carlo program for parton and
  particle production in high-energy hadronic and nuclear collisions},''
  \href{http://dx.doi.org/10.1016/0010-4655(94)90057-4}{{\em Comput. Phys.
  Commun.} {\bfseries 83} (1994) 307},
\href{http://arxiv.org/abs/nucl-th/9502021}{{\ttfamily arXiv:nucl-th/9502021
  [nucl-th]}}.

\bibitem{Miller:2007ri}
M.~L. Miller, K.~Reygers, S.~J. Sanders, and P.~Steinberg, ``{Glauber modeling
  in high energy nuclear collisions},''
  \href{http://dx.doi.org/10.1146/annurev.nucl.57.090506.123020}{{\em Ann. Rev.
  Nucl. Part. Sci.} {\bfseries 57} (2007) 205--243},
\href{http://arxiv.org/abs/nucl-ex/0701025}{{\ttfamily arXiv:nucl-ex/0701025
  [nucl-ex]}}.

\bibitem{Niemi:2015qia}
H.~Niemi, K.~J. Eskola, and R.~Paatelainen, ``{Event-by-event fluctuations in
  perturbative QCD + saturation + hydro model: pinning down QCD matter shear
  viscosity in ultrarelativistic heavy-ion collisions},''
\href{http://arxiv.org/abs/1505.02677}{{\ttfamily arXiv:1505.02677 [hep-ph]}}.

\bibitem{Paatelainen:2012at}
R.~Paatelainen, K.~J. Eskola, H.~Holopainen, and K.~Tuominen, ``{Multiplicities
  and $p_T$ spectra in ultrarelativistic heavy ion collisions from a
  next-to-leading order improved perturbative QCD + saturation + hydrodynamics
  model},'' \href{http://dx.doi.org/10.1103/PhysRevC.87.044904}{{\em Phys.
  Rev.} {\bfseries C87} no.~4, (2013) 044904},
\href{http://arxiv.org/abs/1211.0461}{{\ttfamily arXiv:1211.0461 [hep-ph]}}.

\bibitem{Paatelainen:2013eea}
R.~Paatelainen, K.~J. Eskola, H.~Niemi, and K.~Tuominen, ``{Fluid dynamics with
  saturated minijet initial conditions in ultrarelativistic heavy-ion
  collisions},'' \href{http://dx.doi.org/10.1016/j.physletb.2014.02.018}{{\em
  Phys. Lett.} {\bfseries B731} (2014) 126--130},
\href{http://arxiv.org/abs/1310.3105}{{\ttfamily arXiv:1310.3105 [hep-ph]}}.

\bibitem{Lin:2004en}
Z.-W. Lin, C.~M. Ko, B.-A. Li, B.~Zhang, and S.~Pal, ``{A Multi-phase transport
  model for relativistic heavy ion collisions},''
  \href{http://dx.doi.org/10.1103/PhysRevC.72.064901}{{\em Phys. Rev.}
  {\bfseries C72} (2005) 064901},
\href{http://arxiv.org/abs/nucl-th/0411110}{{\ttfamily arXiv:nucl-th/0411110
  [nucl-th]}}.

\bibitem{Aad:2015lwa}
{\bfseries ATLAS} Collaboration, G.~Aad {\em et~al.}, ``{Measurement of the
  correlation between flow harmonics of different order in lead-lead collisions
  at $\sqrt{s_{NN}}$=2.76 TeV with the ATLAS detector},''
  \href{http://dx.doi.org/10.1103/PhysRevC.92.034903}{{\em Phys. Rev.}
  {\bfseries C92} no.~3, (2015) 034903},
\href{http://arxiv.org/abs/1504.01289}{{\ttfamily arXiv:1504.01289 [hep-ex]}}.

\bibitem{Schukraft:2012ah}
J.~Schukraft, A.~Timmins, and S.~A. Voloshin, ``{Ultra-relativistic nuclear
  collisions: event shape engineering},''
  \href{http://dx.doi.org/10.1016/j.physletb.2013.01.045}{{\em Phys. Lett.}
  {\bfseries B719} (2013) 394--398},
\href{http://arxiv.org/abs/1208.4563}{{\ttfamily arXiv:1208.4563 [nucl-ex]}}.

\bibitem{Petersen:2013vca}
H.~Petersen and B.~Muller, ``{Possibility of event shape selection in
  relativistic heavy ion collisions},''
  \href{http://dx.doi.org/10.1103/PhysRevC.88.044918}{{\em Phys. Rev.}
  {\bfseries C88} no.~4, (2013) 044918},
\href{http://arxiv.org/abs/1305.2735}{{\ttfamily arXiv:1305.2735 [nucl-th]}}.

\bibitem{Eremin:2003qn}
S.~Eremin and S.~Voloshin, ``{Nucleon participants or quark participants?},''
  \href{http://dx.doi.org/10.1103/PhysRevC.67.064905}{{\em Phys. Rev.}
  {\bfseries C67} (2003) 064905},
\href{http://arxiv.org/abs/nucl-th/0302071}{{\ttfamily arXiv:nucl-th/0302071
  [nucl-th]}}.

\bibitem{Miller:2003kd}
M.~Miller and R.~Snellings, ``{Eccentricity fluctuations and its possible
  effect on elliptic flow measurements},''
\href{http://arxiv.org/abs/nucl-ex/0312008}{{\ttfamily arXiv:nucl-ex/0312008
  [nucl-ex]}}.

\bibitem{Adamczyk:2015obl}
{\bfseries STAR} Collaboration, L.~Adamczyk {\em et~al.}, ``{Azimuthal
  anisotropy in U$+$U and Au$+$Au collisions at RHIC},''
  \href{http://dx.doi.org/10.1103/PhysRevLett.115.222301}{{\em Phys. Rev.
  Lett.} {\bfseries 115} no.~22, (2015) 222301},
\href{http://arxiv.org/abs/1505.07812}{{\ttfamily arXiv:1505.07812 [nucl-ex]}}.

\bibitem{Adare:2015bua}
{\bfseries PHENIX} Collaboration, A.~Adare {\em et~al.}, ``{Transverse energy
  production and charged-particle multiplicity at midrapidity in various
  systems from $\sqrt{s_{NN}}=7.7$ to 200 GeV},''
  \href{http://dx.doi.org/10.1103/PhysRevC.93.024901}{{\em Phys. Rev.}
  {\bfseries C93} no.~2, (2016) 024901},
\href{http://arxiv.org/abs/1509.06727}{{\ttfamily arXiv:1509.06727 [nucl-ex]}}.

\bibitem{Loizides:2016djv}
C.~Loizides, ``{Glauber modeling of high-energy nuclear collisions at
  sub-nucleon level},''
\href{http://arxiv.org/abs/1603.07375}{{\ttfamily arXiv:1603.07375 [nucl-ex]}}.

\end{thebibliography}\endgroup

\newpage
\appendix
\section{The ALICE Collaboration}
\label{app:collab}



\begingroup
\small
\begin{flushleft}
J.~Adam$^\textrm{\scriptsize 39}$,
D.~Adamov\'{a}$^\textrm{\scriptsize 85}$,
M.M.~Aggarwal$^\textrm{\scriptsize 89}$,
G.~Aglieri Rinella$^\textrm{\scriptsize 35}$,
M.~Agnello$^\textrm{\scriptsize 112}$\textsuperscript{,}$^\textrm{\scriptsize 31}$,
N.~Agrawal$^\textrm{\scriptsize 48}$,
Z.~Ahammed$^\textrm{\scriptsize 136}$,
S.~Ahmad$^\textrm{\scriptsize 18}$,
S.U.~Ahn$^\textrm{\scriptsize 69}$,
S.~Aiola$^\textrm{\scriptsize 140}$
% ...
\renewcommand\labelenumi{\textsuperscript{\theenumi}~}

\section*{Affiliation notes}
\renewcommand\theenumi{\roman{enumi}}
\begin{Authlist}
\item \Adef{0}Deceased
\item \Adef{idp28841940}{Also at: Georgia State University, Atlanta, Georgia, United States}
\item \Adef{idp29562156}{Also at: Also at Department of Applied Physics, Aligarh Muslim University, Aligarh, India}
\item \Adef{idp29928484}{Also at: M.V. Lomonosov Moscow State University, D.V. Skobeltsyn Institute of Nuclear, Physics, Moscow, Russia}
\\
\end{Authlist}

\section*{Collaboration Institutes}
\renewcommand\theenumi{\arabic{enumi}~}

$^{1}$A.I. Alikhanyan National Science Laboratory (Yerevan Physics Institute) Foundation, Yerevan, Armenia
\\
$^{2}$Benem\'{e}rita Universidad Aut\'{o}noma de Puebla, Puebla, Mexico
\\
$^{3}$Bogolyubov Institute for Theoretical Physics, Kiev, Ukraine
\\
$^{4}$Bose Institute, Department of Physics 
and Centre for Astroparticle Physics and Space Science (CAPSS), Kolkata, India
\\
$^{5}$Budker Institute for Nuclear Physics, Novosibirsk, Russia
\\
$^{6}$California Polytechnic State University, San Luis Obispo, California, United States
\\
$^{7}$Central China Normal University, Wuhan, China
\\
$^{8}$Centre de Calcul de l'IN2P3, Villeurbanne, Lyon, France
\\
$^{9}$Centro de Aplicaciones Tecnol\'{o}gicas y Desarrollo Nuclear (CEADEN), Havana, Cuba
\\
$^{10}$Centro de Investigaciones Energ\'{e}ticas Medioambientales y Tecnol\'{o}gicas (CIEMAT), Madrid, Spain
\\
$^{11}$Centro de Investigaci\'{o}n y de Estudios Avanzados (CINVESTAV), Mexico City and M\'{e}rida, Mexico
\\
$^{12}$Centro Fermi - Museo Storico della Fisica e Centro Studi e Ricerche ``Enrico Fermi', Rome, Italy
\\
$^{13}$Chicago State University, Chicago, Illinois, United States
\\
$^{14}$China Institute of Atomic Energy, Beijing, China
\\
$^{15}$Commissariat \`{a} l'Energie Atomique, IRFU, Saclay, France
\\
$^{16}$COMSATS Institute of Information Technology (CIIT), Islamabad, Pakistan
\\
$^{17}$Departamento de F\'{\i}sica de Part\'{\i}culas and IGFAE, Universidad de Santiago de Compostela, Santiago de Compostela, Spain
\\
$^{18}$Department of Physics, Aligarh Muslim University, Aligarh, India
\\
$^{19}$Department of Physics, Ohio State University, Columbus, Ohio, United States
\\
$^{20}$Department of Physics, Sejong University, Seoul, South Korea
\\
$^{21}$Department of Physics, University of Oslo, Oslo, Norway
\\
$^{22}$Department of Physics and Technology, University of Bergen, Bergen, Norway
\\
$^{23}$Dipartimento di Fisica dell'Universit\`{a} 'La Sapienza'
and Sezione INFN, Rome, Italy
\\
$^{24}$Dipartimento di Fisica dell'Universit\`{a}
and Sezione INFN, Cagliari, Italy
\\
$^{25}$Dipartimento di Fisica dell'Universit\`{a}
and Sezione INFN, Trieste, Italy
\\
$^{26}$Dipartimento di Fisica dell'Universit\`{a}
and Sezione INFN, Turin, Italy
\\
$^{27}$Dipartimento di Fisica e Astronomia dell'Universit\`{a}
and Sezione INFN, Bologna, Italy
\\
$^{28}$Dipartimento di Fisica e Astronomia dell'Universit\`{a}
and Sezione INFN, Catania, Italy
\\
$^{29}$Dipartimento di Fisica e Astronomia dell'Universit\`{a}
and Sezione INFN, Padova, Italy
\\
$^{30}$Dipartimento di Fisica `E.R.~Caianiello' dell'Universit\`{a}
and Gruppo Collegato INFN, Salerno, Italy
\\
$^{31}$Dipartimento DISAT del Politecnico and Sezione INFN, Turin, Italy
\\
$^{32}$Dipartimento di Scienze e Innovazione Tecnologica dell'Universit\`{a} del Piemonte Orientale and INFN Sezione di Torino, Alessandria, Italy
\\
$^{33}$Dipartimento Interateneo di Fisica `M.~Merlin'
and Sezione INFN, Bari, Italy
\\
$^{34}$Division of Experimental High Energy Physics, University of Lund, Lund, Sweden
\\
$^{35}$European Organization for Nuclear Research (CERN), Geneva, Switzerland
\\
$^{36}$Excellence Cluster Universe, Technische Universit\"{a}t M\"{u}nchen, Munich, Germany
\\
$^{37}$Faculty of Engineering, Bergen University College, Bergen, Norway
\\
$^{38}$Faculty of Mathematics, Physics and Informatics, Comenius University, Bratislava, Slovakia
\\
$^{39}$Faculty of Nuclear Sciences and Physical Engineering, Czech Technical University in Prague, Prague, Czech Republic
\\
$^{40}$Faculty of Science, P.J.~\v{S}af\'{a}rik University, Ko\v{s}ice, Slovakia
\\
$^{41}$Faculty of Technology, Buskerud and Vestfold University College, Tonsberg, Norway
\\
$^{42}$Frankfurt Institute for Advanced Studies, Johann Wolfgang Goethe-Universit\"{a}t Frankfurt, Frankfurt, Germany
\\
$^{43}$Gangneung-Wonju National University, Gangneung, South Korea
\\
$^{44}$Gauhati University, Department of Physics, Guwahati, India
\\
$^{45}$Helmholtz-Institut f\"{u}r Strahlen- und Kernphysik, Rheinische Friedrich-Wilhelms-Universit\"{a}t Bonn, Bonn, Germany
\\
$^{46}$Helsinki Institute of Physics (HIP), Helsinki, Finland
\\
$^{47}$Hiroshima University, Hiroshima, Japan
\\
$^{48}$Indian Institute of Technology Bombay (IIT), Mumbai, India
\\
$^{49}$Indian Institute of Technology Indore, Indore, India
\\
$^{50}$Indonesian Institute of Sciences, Jakarta, Indonesia
\\
$^{51}$Inha University, Incheon, South Korea
\\
$^{52}$Institut de Physique Nucl\'eaire d'Orsay (IPNO), Universit\'e Paris-Sud, CNRS-IN2P3, Orsay, France
\\
$^{53}$Institute for Nuclear Research, Academy of Sciences, Moscow, Russia
\\
$^{54}$Institute for Subatomic Physics of Utrecht University, Utrecht, Netherlands
\\
$^{55}$Institute for Theoretical and Experimental Physics, Moscow, Russia
\\
$^{56}$Institute of Experimental Physics, Slovak Academy of Sciences, Ko\v{s}ice, Slovakia
\\
$^{57}$Institute of Physics, Academy of Sciences of the Czech Republic, Prague, Czech Republic
\\
$^{58}$Institute of Physics, Bhubaneswar, India
\\
$^{59}$Institute of Space Science (ISS), Bucharest, Romania
\\
$^{60}$Institut f\"{u}r Informatik, Johann Wolfgang Goethe-Universit\"{a}t Frankfurt, Frankfurt, Germany
\\
$^{61}$Institut f\"{u}r Kernphysik, Johann Wolfgang Goethe-Universit\"{a}t Frankfurt, Frankfurt, Germany
\\
$^{62}$Institut f\"{u}r Kernphysik, Westf\"{a}lische Wilhelms-Universit\"{a}t M\"{u}nster, M\"{u}nster, Germany
\\
$^{63}$Instituto de Ciencias Nucleares, Universidad Nacional Aut\'{o}noma de M\'{e}xico, Mexico City, Mexico
\\
$^{64}$Instituto de F\'{\i}sica, Universidad Nacional Aut\'{o}noma de M\'{e}xico, Mexico City, Mexico
\\
$^{65}$Institut Pluridisciplinaire Hubert Curien (IPHC), Universit\'{e} de Strasbourg, CNRS-IN2P3, Strasbourg, France
\\
$^{66}$iThemba LABS, National Research Foundation, Somerset West, South Africa
\\
$^{67}$Joint Institute for Nuclear Research (JINR), Dubna, Russia
\\
$^{68}$Konkuk University, Seoul, South Korea
\\
$^{69}$Korea Institute of Science and Technology Information, Daejeon, South Korea
\\
$^{70}$KTO Karatay University, Konya, Turkey
\\
$^{71}$Laboratoire de Physique Corpusculaire (LPC), Clermont Universit\'{e}, Universit\'{e} Blaise Pascal, CNRS--IN2P3, Clermont-Ferrand, France
\\
$^{72}$Laboratoire de Physique Subatomique et de Cosmologie, Universit\'{e} Grenoble-Alpes, CNRS-IN2P3, Grenoble, France
\\
$^{73}$Laboratori Nazionali di Frascati, INFN, Frascati, Italy
\\
$^{74}$Laboratori Nazionali di Legnaro, INFN, Legnaro, Italy
\\
$^{75}$Lawrence Berkeley National Laboratory, Berkeley, California, United States
\\
$^{76}$Moscow Engineering Physics Institute, Moscow, Russia
\\
$^{77}$Nagasaki Institute of Applied Science, Nagasaki, Japan
\\
$^{78}$National Centre for Nuclear Studies, Warsaw, Poland
\\
$^{79}$National Institute for Physics and Nuclear Engineering, Bucharest, Romania
\\
$^{80}$National Institute of Science Education and Research, Bhubaneswar, India
\\
$^{81}$National Research Centre Kurchatov Institute, Moscow, Russia
\\
$^{82}$Niels Bohr Institute, University of Copenhagen, Copenhagen, Denmark
\\
$^{83}$Nikhef, Nationaal instituut voor subatomaire fysica, Amsterdam, Netherlands
\\
$^{84}$Nuclear Physics Group, STFC Daresbury Laboratory, Daresbury, United Kingdom
\\
$^{85}$Nuclear Physics Institute, Academy of Sciences of the Czech Republic, \v{R}e\v{z} u Prahy, Czech Republic
\\
$^{86}$Oak Ridge National Laboratory, Oak Ridge, Tennessee, United States
\\
$^{87}$Petersburg Nuclear Physics Institute, Gatchina, Russia
\\
$^{88}$Physics Department, Creighton University, Omaha, Nebraska, United States
\\
$^{89}$Physics Department, Panjab University, Chandigarh, India
\\
$^{90}$Physics Department, University of Athens, Athens, Greece
\\
$^{91}$Physics Department, University of Cape Town, Cape Town, South Africa
\\
$^{92}$Physics Department, University of Jammu, Jammu, India
\\
$^{93}$Physics Department, University of Rajasthan, Jaipur, India
\\
$^{94}$Physikalisches Institut, Eberhard Karls Universit\"{a}t T\"{u}bingen, T\"{u}bingen, Germany
\\
$^{95}$Physikalisches Institut, Ruprecht-Karls-Universit\"{a}t Heidelberg, Heidelberg, Germany
\\
$^{96}$Physik Department, Technische Universit\"{a}t M\"{u}nchen, Munich, Germany
\\
$^{97}$Purdue University, West Lafayette, Indiana, United States
\\
$^{98}$Pusan National University, Pusan, South Korea
\\
$^{99}$Research Division and ExtreMe Matter Institute EMMI, GSI Helmholtzzentrum f\"ur Schwerionenforschung, Darmstadt, Germany
\\
$^{100}$Rudjer Bo\v{s}kovi\'{c} Institute, Zagreb, Croatia
\\
$^{101}$Russian Federal Nuclear Center (VNIIEF), Sarov, Russia
\\
$^{102}$Saha Institute of Nuclear Physics, Kolkata, India
\\
$^{103}$School of Physics and Astronomy, University of Birmingham, Birmingham, United Kingdom
\\
$^{104}$Secci\'{o}n F\'{\i}sica, Departamento de Ciencias, Pontificia Universidad Cat\'{o}lica del Per\'{u}, Lima, Peru
\\
$^{105}$Sezione INFN, Bari, Italy
\\
$^{106}$Sezione INFN, Bologna, Italy
\\
$^{107}$Sezione INFN, Cagliari, Italy
\\
$^{108}$Sezione INFN, Catania, Italy
\\
$^{109}$Sezione INFN, Padova, Italy
\\
$^{110}$Sezione INFN, Rome, Italy
\\
$^{111}$Sezione INFN, Trieste, Italy
\\
$^{112}$Sezione INFN, Turin, Italy
\\
$^{113}$SSC IHEP of NRC Kurchatov institute, Protvino, Russia
\\
$^{114}$Stefan Meyer Institut f\"{u}r Subatomare Physik (SMI), Vienna, Austria
\\
$^{115}$SUBATECH, Ecole des Mines de Nantes, Universit\'{e} de Nantes, CNRS-IN2P3, Nantes, France
\\
$^{116}$Suranaree University of Technology, Nakhon Ratchasima, Thailand
\\
$^{117}$Technical University of Ko\v{s}ice, Ko\v{s}ice, Slovakia
\\
$^{118}$Technical University of Split FESB, Split, Croatia
\\
$^{119}$The Henryk Niewodniczanski Institute of Nuclear Physics, Polish Academy of Sciences, Cracow, Poland
\\
$^{120}$The University of Texas at Austin, Physics Department, Austin, Texas, United States
\\
$^{121}$Universidad Aut\'{o}noma de Sinaloa, Culiac\'{a}n, Mexico
\\
$^{122}$Universidade de S\~{a}o Paulo (USP), S\~{a}o Paulo, Brazil
\\
$^{123}$Universidade Estadual de Campinas (UNICAMP), Campinas, Brazil
\\
$^{124}$Universidade Federal do ABC, Santo Andre, Brazil
\\
$^{125}$University of Houston, Houston, Texas, United States
\\
$^{126}$University of Jyv\"{a}skyl\"{a}, Jyv\"{a}skyl\"{a}, Finland
\\
$^{127}$University of Liverpool, Liverpool, United Kingdom
\\
$^{128}$University of Tennessee, Knoxville, Tennessee, United States
\\
$^{129}$University of the Witwatersrand, Johannesburg, South Africa
\\
$^{130}$University of Tokyo, Tokyo, Japan
\\
$^{131}$University of Tsukuba, Tsukuba, Japan
\\
$^{132}$University of Zagreb, Zagreb, Croatia
\\
$^{133}$Universit\'{e} de Lyon, Universit\'{e} Lyon 1, CNRS/IN2P3, IPN-Lyon, Villeurbanne, Lyon, France
\\
$^{134}$Universit\`{a} di Brescia, Brescia, Italy
\\
$^{135}$V.~Fock Institute for Physics, St. Petersburg State University, St. Petersburg, Russia
\\
$^{136}$Variable Energy Cyclotron Centre, Kolkata, India
\\
$^{137}$Warsaw University of Technology, Warsaw, Poland
\\
$^{138}$Wayne State University, Detroit, Michigan, United States
\\
$^{139}$Wigner Research Centre for Physics, Hungarian Academy of Sciences, Budapest, Hungary
\\
$^{140}$Yale University, New Haven, Connecticut, United States
\\
$^{141}$Yonsei University, Seoul, South Korea
\\
$^{142}$Zentrum f\"{u}r Technologietransfer und Telekommunikation (ZTT), Fachhochschule Worms, Worms, Germany
\endgroup

\end{document}